%% file: main.tex
\documentclass[sigconf, 9pt, nonacm]{acmart}

\renewcommand\footnotetextcopyrightpermission[1]{}

\usepackage{makecell}
\usepackage{booktabs,tabularx,array}
\newcolumntype{Y}{>{\raggedright\arraybackslash}X}

\newcolumntype{N}[1]{>{\raggedright\arraybackslash}p{#1}}

\usepackage{amsmath,amssymb,amsfonts}
\usepackage{algorithmic}
\usepackage{graphicx}
\usepackage{textcomp}
\usepackage{xcolor}
\usepackage{adjustbox}
\usepackage{multirow}
\usepackage{pifont}
\usepackage{makecell} 
\usepackage[most]{tcolorbox}
\usepackage{enumitem}
\usepackage{xspace}
\usepackage[table]{xcolor} 
\newcommand{\baselinecell}[1]{\cellcolor{gray!10}{#1}}
\newcommand{\neutralcell}[1]{\cellcolor{yellow!20}{#1}}
\newcommand{\tealcell}[1]{\cellcolor{green!20}{#1}}     
\newcommand{\purplecell}[1]{\cellcolor{green!45}{\bfseries #1}}

\usepackage{booktabs}
\usepackage{siunitx}
\usepackage[table]{xcolor} 
\usepackage{tikz}          
\usepackage{tcolorbox}
\tcbuselibrary{listings}
\newtcolorbox{promptbox}[2][]{%
  colback=blue!5,        
  colframe=blue!60,      
  colbacktitle=blue!20,  
  coltitle=blue!50!black,        
  fonttitle=\bfseries,   
  fontupper=\ttfamily\small,   
  boxrule=0.6pt,         
  arc=2mm,               
  top=1mm, bottom=1mm,
  left=1mm, right=1mm,
  title=#2,#1
}

\begin{document}

\title{\textit{NetDeTox}: Adversarial and Efficient Evasion of Hardware-Security GNNs via RL-LLM Orchestration\\
}

\author[Zeng~Wang, Minghao~Shao, Akashdeep~Saha, Ramesh~Karri, Johann~Knechtel, Muhammad~Shafique, Ozgur~Sinanoglu]{%
Zeng~Wang$^\dagger$$^\S$,
Minghao~Shao$^\dagger$$^\ddagger$$^\S$,
Akashdeep~Saha$^\ddagger$,
Ramesh~Karri$^\dagger$,
Johann~Knechtel$^\ddagger$,
Muhammad~Shafique$^\ddagger$,
Ozgur~Sinanoglu$^\ddagger$\\[2pt]
$^\dagger$NYU Tandon School of Engineering, New York, USA\\
$^\ddagger$NYU Abu Dhabi, Abu Dhabi, UAE\\[2pt]
\text{\{zw3464, shao.minghao, as19360, rkarri, johann, muhammad.shafique, ozgursin\}@nyu.edu}\\[2pt]
}

\begin{abstract}
\textit{Graph neural network}s (\textit{GNN}s) have shown promise in hardware security 
by learning structural motifs from netlist graphs. However, this reliance 
on motifs makes GNNs vulnerable to adversarial netlist rewrites; even 
small-scale edits can mislead GNN predictions.
Existing adversarial approaches, ranging from synthesis-recipe perturbations to gate transformations, come with high design overheads.
We present \textbf{\textit{NetDeTox}}, an automated end-to-end framework that orchestrates \textit{large language models} (\textit{LLM}s) with \textit{reinforcement learning} (\textit{RL}) in a systematic manner, enabling focused local rewriting.
The RL agent identifies netlist components critical for GNN-based reasoning, while the LLM devises rewriting plans to diversify motifs that preserve functionality.
Iterative feedback between the RL and LLM stages refines adversarial rewritings to limit overheads.
Compared to the SOTA work AttackGNN,
\textit{NetDeTox} successfully degrades the effectiveness of all security schemes with fewer rewrites and substantially lower area overheads (reductions of 54.50\% for GNN-RE, 25.44\%
		for GNN4IP, and 41.04\% for OMLA, respectively).
For GNN4IP, ours can even optimize/reduce the original benchmarks' area, in particular for larger circuits,
demonstrating the practicality and scalability of \textit{NetDeTox}.
\end{abstract}

\maketitle
\begingroup
\renewcommand\thefootnote{$\S$}%
\footnotetext{Authors contributed equally to this research.}%
\endgroup

\input{text/SectionI-Introduction}

\input{text/SectionII-Background}

\input{text/SectionIII-1ThreatModel}
\input{text/SectionIII-Methodology}

\input{text/SectionIV-Experiment_Set}

\input{text/SectionV-1_OMLA-case}

\input{text/SectionV-II_GNN4IP-case}
\input{text/SectionV-III_GNNRE}
\input{text/SectionVI-Discussion}
\input{text/SectionVII-Conclusion}

\bibliographystyle{ACM-Reference-Format}
\bibliography{Reference}

\end{document}

%% file: text/SectionI-Introduction.tex
\section{Introduction}

Graph neural networks (GNNs) have been widely adopted in hardware security, including IP piracy detection~\cite{yasaei2021gnn4ip}, reverse engineering~\cite{alrahis2021gnn}, logic locking attacks~\cite{alrahis2021omla},
      and hardware Trojan detection~\cite{lashen2023trojansaint}. However, recent work shows that GNN-based methods are vulnerable to adversarial netlist transformations using reinforcement learning
      (RL)~\cite{gohil2024attackgnn} to rewrite whole designs or large language models (LLMs)~\cite{gohil2024llmpirate} to introduce more various localized transformations. While these attacks can evade GNN analysis,
      they incur significant design overheads, creating an undesirable trade-off.

While LLMs excel at hardware design tasks such as Verilog generation~\cite{wang2024llms, thakur2024verigen, roy2025veritas}, assertion generation~\cite{kande2023llm, yan2025assertllm}, and testbench
synthesis~\cite{bhandari2024llm, qiu2024autobench}, and have been applied to hardware security analysis~\cite{wang2025vericontaminated, wang2025verileaky, wang2025salad}, their ability to manipulate gate-level netlists
remains limited. The complex semantics and structural features of netlists pose significant challenges for LLM-based transformation~\cite{zhang2025rebert, fang2025geneda}. Recent planning-based code generation approaches~\cite{huang2024understanding, bairi2024codeplan} show that chain-of-edits methods can effectively handle large codebases through iterative, context-aware transformations.

\begin{figure}[!t]
    \centering
    \includegraphics[width=1.0\columnwidth]{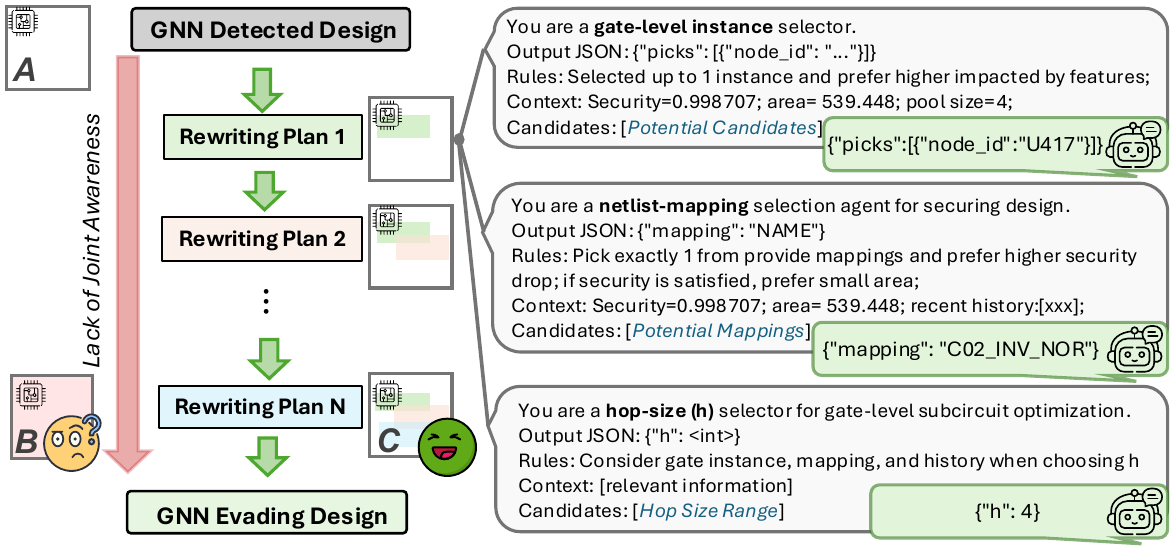} 
    \caption{Demonstration of \textit{NetDeTox} for netlist rewriting. Direct rewriting of a GNN detected design \textit{A}, without joint security–area awareness, produces a suboptimal design \textit{B}. In contrast,
	    \textit{NetDeTox} employs context-aware iterative planning with coordinated gate selection, subnetlist mapping, and hop-size determination, all to achieve an efficient design \textit{C}.}
    \label{fig:motivation}
    \vspace{-2mm}
\end{figure}

Drawing on this, we approach evasion through iterative netlist rewriting, generating a sequence of targeted transformations that adversarially alter GNN-detected graph structures while optimizing for area and preserving functionality.
This approach requires addressing a key challenge: how to effectively reason about netlist transformations. Pure LLM-based methods struggle with gate-level representations~\cite{gohil2024llmpirate, fang2025nettag}, while RL excels at structural netlist processing~\cite{chowdhury2024retrieval, gohil2024attackgnn}. We bridge this gap through a hybrid architecture where RL identifies critical transformation locations while LLMs provide context-aware reasoning for hardware security applications.

We propose \textit{NetDeTox}, a novel framework that combines RL-based gate selection with LLM-driven rewriting to progressively evade GNN-based security tools. Rather than processing entire designs through LLMs,
   \textit{NetDeTox} focuses on targeted gate pools identified by RL. As shown in Fig.~\ref{fig:motivation}, our framework employs an LLM-in-the-loop process that iteratively generates rewriting plans by coordinating
   gate selection, subnetlist mapping, and hop-size determination. This coordinated planning enables hardware-aware transformations that jointly optimize for security evasion and area efficiency, achieving evasion with
   minimal overheads.

   Our main contributions are:

\begin{itemize}[leftmargin=*, itemsep=0pt, topsep=2pt]

    \item We introduce \textit {NetDeTox}, an RL-assisted, LLM-guided framework for subnetlist rewriting that evades GNN-based analyses while respecting area and implementation constraints.

    \item We demonstrate consistent shortcomings of robust GNN tools (OMLA, GNN4IP, GNN-RE) across six diverse LLM backends, indicating robustness beyond a single model family.
    
    \item We present a detailed evaluation, including comparative case studies and ablation studies (e.g., for the impact of planning order and LLM strategy selection on evasion efficacy and cost).

\end{itemize}

%% file: text/SectionII-Background.tex
\section{Background And Related Works}

\begin{figure*}[!t]
    \centering
    \includegraphics[width=1.0\linewidth]{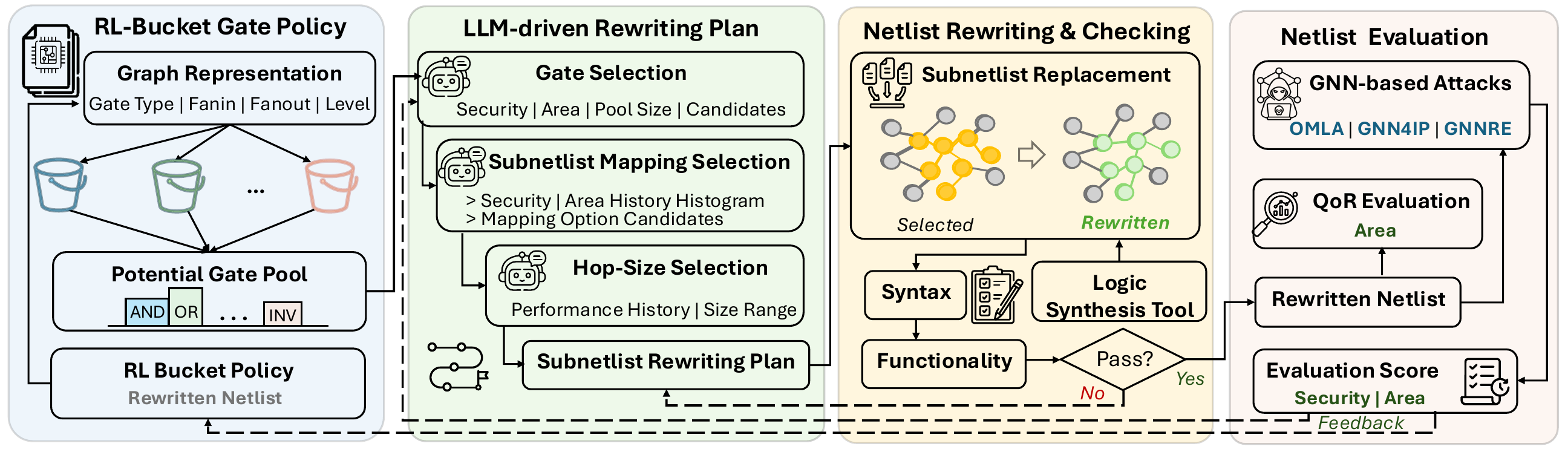} 
    \caption{\textit{NetDeTox} framework overview.}
    \label{fig:framework}
    \vspace{-5mm}
\end{figure*}

\subsection{GNNs for Hardware Security}

GNNs learn graph representations by aggregating information from node neighborhoods. Circuit netlists naturally map to graphs with gates as nodes and nets as edges, making GNNs well-suited for hardware security analysis~\cite{el2024graph}.

GNN-based methods have been applied to both defensive and offensive hardware security tasks. On the defensive side, GNN4TJ~\cite{yasaei2021gnn4tj} targets hardware Trojan (HT) detection in third-party IP by flagging
HT-related structures, while GNN4IP~\cite{yasaei2021gnn4ip} addresses IP piracy by embedding circuit pairs and measuring their similarity (e.g., cosine distance) to detect unauthorized copies. On the offensive side,
OMLA~\cite{alrahis2021omla} attacks synthesized logic-locked designs by exploiting structural context around key-controlled gates to predict key bits with high accuracy, and GNNRE~\cite{alrahis2021gnn} performs reverse
engineering through node-level classification, assigning gates to their original functional modules with approximately 98.8\% accuracy on standard benchmarks. These applications span diverse GNN architectures and
settings, including node classification, graph classification, and graph similarity learning.


\subsection{Adversarial Netlist Generation}

Recent work has explored generating adversarial netlists to evade GNN-based hardware security tools. AttackGNN~\cite{gohil2024attackgnn} employs RL to discover synthesis recipes that systematically rewrite entire
designs, successfully undermining GNN4IP, GNNRE, OMLA, and other GNN-based tools. However, it couples rewriting with per-cycle RL reward updates, increasing computational complexity and runtime.
LLMPirate~\cite{gohil2024llmpirate} leverages LLMs for localized gate transformations to evade GNN4IP, but treats the LLM as a template-driven substitution engine with limited circuit-context awareness. Both approaches
prioritize evasion effectiveness over design quality, resulting in significant area and timing overheads. While recent work~\cite{zhao2024adversarial} addresses area reduction during circuit transformation, it lacks
evaluation against real GNN-based security tools. This motivates our work, achieving effective GNN evasion while maintaining design quality through context-aware, targeted transformations.

%% file: text/SectionIII-1ThreatModel.tex
\section{Threat Model}

We consider a standard black-box adversary~\cite{gohil2024attackgnn, gohil2024llmpirate} who seeks to evade GNN-based security tools, namely OMLA, GNN4IP, and GNNRE.
{Each tool returns a real-valued security score: GNN4IP provides a similarity ranging from $-1$ to $1$, while OMLA and GNNRE return classification confidence scores ranging from $0$ to $1$.}
The adversary operates post-deployment with access only
to the detector's input-output interface, receiving scores for any netlist provided.
The adversary has no knowledge of internal parameters, training data, or model architecture. The adversary cannot alter or retrain these GNN-based security tools.

Our framework requires no knowledge of specific GNN vulnerabilities—it operates purely through black-box queries. Netlist transformations must preserve functional equivalence and respect standard design rules.
Our evaluation demonstrates transferability across multiple GNN architectures and LLM backends. The adversary's goal is generating functionally equivalent netlists that evade detection while maintaining design quality.

%% file: text/SectionIII-Methodology.tex
\section{Our Framework: \textit{NetDeTox}}

Prior approaches~\cite{gohil2024attackgnn, gohil2024llmpirate} operate at whole-netlist scale, possibly inflating silicon area. Moreover, presenting entire netlists to LLMs poses practical challenges: (i) large inputs
incur substantial token costs, and (ii) raw gate-level abstractions offer limited semantic clarity for effective reasoning. To address these challenges, we propose \textit{NetDeTox} (Fig.~\ref{fig:framework}), which
employs an RL-guided policy to assemble gate pools and an LLM to generate context-aware rewriting plans that localize edits while preserving functional and implementation constraints. This targeted pipeline achieves
comparable GNN evasion with significantly lower area overheads.


\subsection{RL-Guided Gate Pooling}
We represent the netlist as a directed graph, with gates as nodes and wires as edges. Each node is annotated with features like gate type, fan-in, fan-out, and logic level, which are used to bin nodes and steer
diversity. An RL policy samples from these bins to form candidate gate pools, directing selection toward high-impact regions rather than repeatedly visiting similar structures.
{More specifically, after each rewrite, the policy updates its bin preferences using a lightweight reward $R = \alpha\,\Delta\text{Security} - \beta\,\Delta\text{Area}$.
Each step consists of (1) constructing bins, (2) selecting a bin via the RL policy, (3) sampling gates from that bin, and (4) updating both the policy and the bins.}
This produces a compact, non-redundant set of RL-selected targets for
subsequent LLM-driven rewriting and is critical to downstream evasion performance.

\subsection{LLM Planning for Subnetlist Rewriting}

Given the RL-selected pools, the LLM orchestrates subnetlist rewrites by setting the strategy and neighbourhood as follows.

\noindent
\textbf{Gate Selection:} 
At each iteration, the LLM receives a pool sampled from the updated bin for that round, which reflects prior edits, security/area feedback, allowed gate types, and structural rules. The LLM selects $N$ target gates from the pool.

\noindent
\textbf{Subnetlist Mapping:}
The LLM chooses from pre-defined mapping options (Tab.~\ref{tab:gate_combos}) to replace logic with distinct primitive compositions, preserving functionality and increasing structural diversity. A prioritized histogram
of historical area and security values guides the choices, so that different regions adopt tailored motifs with limited overlaps with GNN-learned motifs.

\input{tabs/gate_mapping_table}

\noindent
\textbf{Hop-Size Selection:}
During subnetlist construction, the LLM selects a hop size $h$ of logic levels around the chosen gates to include. Larger $h$ enables extensive structural rewriting while smaller $h$ localizes rewriting. The choice is
based on some pre-defined range and informed by prior outcomes.

\noindent
\textbf{Transformation Planning Order:}
We schedule decisions in a logical, policy-driven progression: gate selection, mapping choice, then $h$ determination, enabling the LLM to adapt its reasoning to the evolving circuit context (see Sec.~\ref {ordering}).

\subsection{Subnetlist Rewriting and Validation}

\noindent
\textbf{Rewriting:}
Using the LLM-guided transformation plan, the framework extracts and rewrites the subnetlist defined by the selected gates and hop size $h$. The subnetlist is isolated for local rewriting with LLM-chosen mappings, and the rewritten subnetlist replaces the original one in the  design.

\noindent
\textbf{Syntax/Functional Validation:}
The plan is rejected if the subnetlist introduces syntax errors, connectivity issues, or functional mismatches (Fig.~\ref{fig:framework}). LLM revises its plan, ensuring only valid, functional designs advance.

\subsection{Putting It All Together}
Shown in Fig.~\ref{fig:framework}, \textit{NetDeTox} operates in an iterative loop: node features drive bin segregation, RL samples candidate gates, LLM selects targets, chooses mappings, and sets
$h$. The resulting subnetlist is extracted, rewritten, and inserted back. The rewritings are validated. Next, the modified netlist is fed to the target GNN-based tools to evaluate its security. Finally, the updated
security and area metrics are fed back into the next LLM prompt and the RL bin policy, closing the loop and progressively improving evasion.
\label{net_eval}




%

%% file: tabs/gate_mapping_table.tex
\begin{table}[t]
  \centering
  \scriptsize
  \caption{Predefined Subnetlist Mapping Options.}
  \vspace{-2mm}
  \label{tab:gate_combos}
  \begin{tabular}{p{0.35cm}|p{2.8cm} p{0.35cm}|p{3.7cm}}
    \toprule
    \textbf{Map.} & \textbf{Gate Composition} & \textbf{Map.} & \textbf{Gate Composition} \\
    \midrule
    C01 & INV, NAND, BUF        & C11 & INV, NOR, XOR, BUF \\
    C02 & INV, NOR, BUF         & C12 & INV, NOR, XNOR, BUF \\
    C03 & INV, NAND, LOGIC, BUF & C13 & INV, AND, OR, BUF \\
    C04 & INV, NAND, AND, BUF   & C14 & INV, AND, OR, LOGIC, BUF \\
    C05 & INV, NAND, OR, BUF    & C15 & INV, AND, OR, XOR, BUF \\
    C06 & INV, NAND, XOR, BUF   & C16 & INV, AND, OR, XNOR, BUF \\
    C07 & INV, NAND, XNOR, BUF  & C17 & INV, NAND, LOGIC, XOR, BUF \\
    C08 & INV, NOR, LOGIC, BUF  & C18 & INV, NAND, LOGIC, XNOR, BUF \\
    C09 & INV, NOR, AND, BUF    & C19 & INV, NOR, LOGIC, XOR, BUF \\
    C10 & INV, NOR, OR, BUF     & C20 & INV, NOR, LOGIC, XNOR, BUF \\
    \bottomrule
  \end{tabular}
  \vspace{-7mm}
\end{table}

%% file: text/SectionIV-Experiment_Set.tex
\section{Experimental Investigation}





\subsection{Setup}

\noindent
\textbf{LLM Planning.}  
The LLM generates rewriting plans sequentially: first, it selects $N=5$ target gates, then it chooses one of 20 predefined mapping options (Table~\ref{tab:gate_combos}), and finally it sets a hop size $h \in [1,20]$ for
the rewriting of the neighbourhood. All selected gates share the same mapping and $h$ to reduce prompt overhead. Prompts incorporate prior \textit{security} and \textit{area} metrics for planning.

\noindent
\textbf{RL-Guided Binning.}  
The proposed framework partitions nodes/gates into up to 24 bins. In each round, 20 candidates are sampled via a softmax over bin scores, with per-bin quotas to maintain structural diversity. RL updates use
\textit{reinforce} policy gradients~\cite{williams1992simple}, with reward defined as a weighted sum of security score reduction ($\alpha=1.5$) and area-overhead reduction ($\beta=0.5$). The weights are empirically determined to adjust bin preferences across iterations.

  
\noindent
\textbf{Netlist Evaluation.}
Subnetlist rewriting is performed in ABC~\cite{mishchenko2007abc} using \{\texttt{rewrite -z}, \texttt{refactor -z}, \texttt{resub -z}\} as specified by LLM-selected mappings. Security is assessed by querying pre-trained
models (OMLA, GNN4IP, GNN-RE) for detection/classification scores. Area is evaluated via Yosys~\cite{wolf2013yosys} synthesis for NanGate45~\cite{nangate45}. Following AttackGNN and LLMPirate, for comparison, we consider only area overheads.

\noindent
\textbf{LLM Selection}
We benchmarked six LLM backends to avoid bias: \texttt{LlaMA-4-Maverick-17B}, \texttt{GPT-4o-mini}, \texttt{GPT-5}, \texttt{Qwen-3-235B}, \\ \texttt{DeepSeek-V3}, and \texttt{Gemini-2.5-Flash}. All models used a 2,048 token budget (reasoning included) with temperature $= 0.8$. Prompts enforce JSON outputs for reproducibility.



\noindent
\textbf{Scope of Experiments}
We evaluated \textit{NetDeTox} on SOTA GNN-based tools: OMLA, GNN4IP, and GNN-RE, all on the same benchmarks synthesis flows, and backend LLMs as in~\cite{gohil2024attackgnn}.
We also evaluated \textit{NetDeTox} on GNN4IP, the tool used in \cite{gohil2024llmpirate}.
Our analysis shows not only strong evasion performance but also consistent cost savings and scalability.
Finally, we conduct various ablation studies, confirming the practical relevance of our proposed framework.

%% file: text/SectionV-1_OMLA-case.tex
\subsection{Case Study 1: \textbf{OMLA} and AttackGNN}

\begin{figure}[htbp]
    \centering
    \includegraphics[width=1.0\linewidth]{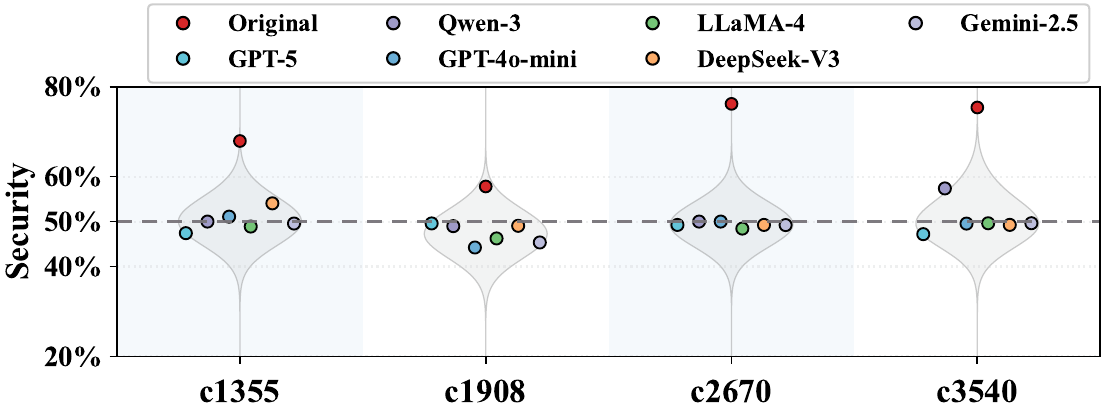} 
    \caption{Security of \textit{NetDeTox} netlists against OMLA. Original refers to the baseline design. Scores $\approx 50\%$ indicate successful evasion; the original's score $> 50\%$ shows high baseline vulnerability to OMLA.}
    \label{fig:omla}
\end{figure}

\noindent\textbf{Security Evaluation:} 
For OMLA, successful evasion is indicated by key accuracy around 50\% (random guessing), meaning the attack cannot recover the secret key. \textit{NetDeTox} achieves this threshold in 23/24 benchmark-backend pairs
(Fig.~\ref{fig:omla}), with only Qwen-3 on \texttt{c3540} showing marginal underperformance.
{Compared to AttackGNN, which pushes OMLA scores between $45\%\!$ and $55\%$, \textit{NetDeTox} drives them more consistently toward the $50\%$ region.}

\noindent
\textbf{Overhead Analysis:} 
From Tab.~\ref{tab:overhead_omla}, \textit{NetDeTox} achieves more favorable area-security trade-offs than AttackGNN. For example, on \texttt{c1908}, it achieves as little as 24.79\% area overheads, whereas AttackGNN
incurs 80.28\% overheads.
Reduced overheads scale well with larger designs, which offer more subnetlist optimization opportunities. Critically, targeted subnetlist rewriting avoids the potentially exponential area growth of global rewriting.


\input{figs/omla_area}

%% file: figs/omla_area.tex

\begin{table}[!t]
\centering
\caption{Area overhead [\%]: AttackGNN vs \textit{NetDeTox} on OMLA. \textbf{Abbreviations:} AG = AttackGNN, G4oM = GPT-4o-mini, L4 = LLaMA-4 Maverick 17B, DS-V3 = Deepseek-V3, Q3 = Qwen-3 235B, G2.5 = Gemini-2.5 Flash,
	G5 = GPT-5.}
\label{tab:overhead_omla}
\setlength{\tabcolsep}{6pt}
\renewcommand{\arraystretch}{1.0}
\scriptsize
\begin{tabular}{l c|cccccc}
\toprule
\textbf{Case} & \textbf{AG} & \textbf{G4oM} & \textbf{L4} & \textbf{DS-V3} & \textbf{Q3} & \textbf{G2.5} & \textbf{G5} \\
\midrule
c1355 & \baselinecell{121.93}
& \neutralcell{138.88}
& \tealcell{116.64}
& \neutralcell{144.48}
& \tealcell{106.61}
& \tealcell{111.59}
& \purplecell{72.55} \\
c1908 & \baselinecell{80.28}
& \tealcell{44.51}
& \purplecell{24.79}
& \tealcell{43.54}
& \tealcell{42.01}
& \tealcell{32.50}
& \tealcell{59.51} \\
c2670 & \baselinecell{77.23}
& \tealcell{39.69}
& \purplecell{9.24}
& \tealcell{11.12}
& \tealcell{75.89}
& \tealcell{69.87}
& \tealcell{10.15} \\
c3540 & \baselinecell{50.57}
& \purplecell{17.57}
& \tealcell{28.96}
& \tealcell{34.44}
& \tealcell{34.04}
& \tealcell{35.00}
& \tealcell{23.61} \\
\bottomrule
\end{tabular}

\footnotesize
\noindent
\textit{Overhead Status: }
\colorbox{gray!20}{\rule{0pt}{4pt}\rule{8pt}{0pt}} baseline\,
\colorbox{yellow!20}{\rule{0pt}{4pt}\rule{8pt}{0pt}} worse\,
\colorbox{green!20}{\rule{0pt}{4pt}\rule{8pt}{0pt}} better\,
\colorbox{green!45}{\rule{0pt}{4pt}\rule{8pt}{0pt}} best

\end{table}

%% file: text/SectionV-II_GNN4IP-case.tex
\input{figs/gnnip_area}

\subsection{Case Study 2: \textbf{GNN4IP} and AttackGNN}
\label{sec:case_GNN4IP}

\begin{figure*}[!t]
    \centering
    \includegraphics[width=1.0\linewidth]{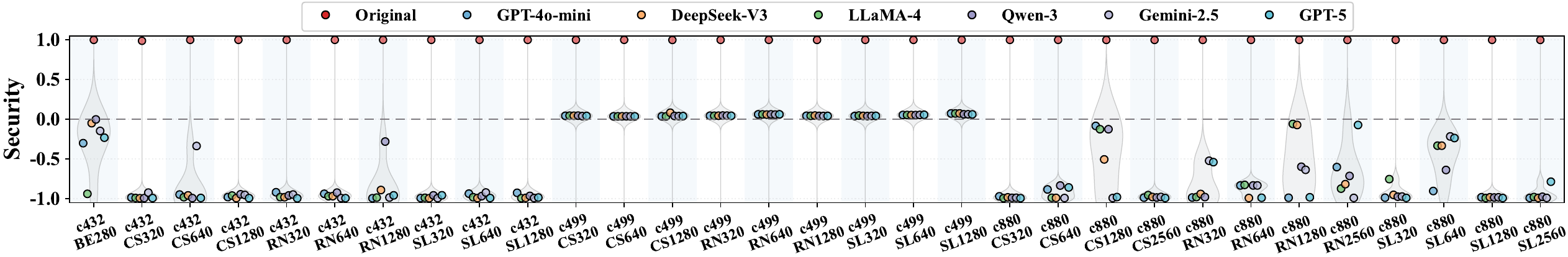} 
    \caption{Security of \textit{NetDeTox} netlists against GNN4IP. Original refers to the baseline design. Scores $< 0$ indicate successful evasion; the original's score $\approx 1$ shows high baseline vulnerability to GNN4IP.}
    \label{fig:gnnip}
\end{figure*}

\noindent\textbf{Security Evaluation:} 
For GNN4IP, successful evasion requires similarity scores $\leq 0$, indicating designs appear unrelated. \textit{NetDeTox} achieves this threshold in 90.3\% of cases (168/186 model-benchmark pairs, Fig.~\ref{fig:gnnip}),
    with LLaMA-4 and DeepSeek-V3 consistently reaching $\leq 0$ across synthesis flows. Interestingly, some benchmarks like \texttt{c499} achieve near-zero scores across all backends, demonstrating successful evasion
    even on familiar training circuits.
{In comparison, AttackGNN leaves scores between $0$ and $-1$, while \textit{NetDeTox} pushes almost all designs toward $-1$.}

\noindent
\textbf{Overhead Analysis:} 
Notably, {41.9\%}
of all cases show area reductions (Tab.~\ref{tab:overhead_extended}). For instance, \texttt{c880-RN320} achieves a security score of $-0.991$ with -2.51\% area overhead. This occurs because localized subnetlist rewrites
enable better area control during structural changes, allowing us to discover more efficient implementations while evading detection. Even when area increases, our results remain substantially better than AttackGNN's overhead.
This demonstrates that security and efficiency are not necessarily conflicting objectives: \textit{NetDeTox} can harden designs while reducing area.


%% file: figs/gnnip_area.tex
\begin{table}[htbp]
\centering
\caption{Area overhead: AttackGNN vs \textit{NetDeTox} on GNN4IP.}
\label{tab:overhead_extended}
\setlength{\tabcolsep}{5pt}
\renewcommand{\arraystretch}{1.0}
\scriptsize
\begin{tabular}{l c|cccccc}
\toprule
\textbf{Case} & \textbf{AG} & \textbf{G4oM} & \textbf{L4} & \textbf{DS-V3} & \textbf{Q3} & \textbf{G2.5} & \textbf{G5} \\
\midrule
c432-BE280  & \baselinecell{48.55}
& \tealcell{32.51}
& \purplecell{16.23}
& \tealcell{27.25}
& \tealcell{33.64}
& \tealcell{21.70}
& \tealcell{27.35} \\

c432-CS320  & \baselinecell{99.69}
& \tealcell{76.55}
& \purplecell{25.31}
& \tealcell{46.74}
& \neutralcell{129.04}
& \tealcell{96.43}
& \neutralcell{228.88} \\

c432-CS640  & \baselinecell{114.04}
& \purplecell{16.02}
& \tealcell{75.44}
& \tealcell{53.22}
& \tealcell{113.45}
& \tealcell{115.20}
& \neutralcell{130.76} \\

c432-CS1280 & \baselinecell{121.80}
& \purplecell{8.29}
& \tealcell{88.31}
& \tealcell{78.12}
& \tealcell{115.72}
& \neutralcell{149.61}
& \neutralcell{154.90} \\

c432-RN320  & \baselinecell{71.43}
& \neutralcell{72.17}
& \tealcell{25.60}
& \purplecell{20.68}
& \neutralcell{97.47}
& \tealcell{56.40}
& \neutralcell{169.20} \\

c432-RN640  & \baselinecell{105.76}
& \neutralcell{116.08}
& \neutralcell{155.43}
& \neutralcell{108.31}
& \purplecell{90.91}
& \tealcell{95.01}
& \neutralcell{121.62} \\

c432-RN1280 & \baselinecell{115.87}
& \neutralcell{137.00}
& \purplecell{8.86}
& \neutralcell{141.37}
& \neutralcell{131.97}
& \tealcell{115.64}
& \tealcell{108.09} \\

c432-SL320  & \baselinecell{96.79}
& \neutralcell{114.89}
& \tealcell{40.29}
& \purplecell{35.91}
& \neutralcell{155.77}
& \tealcell{79.71}
& \tealcell{73.43} \\

c432-SL640  & \baselinecell{125.89}
& \neutralcell{199.64}
& \tealcell{87.95}
& \purplecell{22.43}
& \neutralcell{184.61}
& \neutralcell{183.77}
& \neutralcell{126.61} \\

c432-SL1280 & \baselinecell{127.95}
& \neutralcell{184.01}
& \purplecell{92.00}
& \tealcell{97.25}
& \tealcell{127.46}
& \neutralcell{195.07}
& \neutralcell{176.66} \\

c499-CS320  & \baselinecell{81.56}
& \purplecell{54.87}
& \tealcell{81.03}
& \neutralcell{92.28}
& \tealcell{85.38}
& \tealcell{87.03}
& \neutralcell{122.41} \\

c499-CS640  & \baselinecell{87.36}
& \tealcell{87.55}
& \neutralcell{102.94}
& \purplecell{84.36}
& \neutralcell{126.82}
& \neutralcell{101.92}
& \tealcell{87.23} \\

c499-CS1280 & \baselinecell{93.49}
& \tealcell{92.55}
& \tealcell{93.10}
& \purplecell{27.71}
& \neutralcell{107.54}
& \tealcell{72.49}
& \neutralcell{110.95} \\

c499-RN320  & \baselinecell{78.49}
& \neutralcell{82.86}
& \purplecell{75.37}
& \tealcell{84.20}
& \neutralcell{98.22}
& \tealcell{90.65}
& \neutralcell{128.56} \\

c499-RN640  & \baselinecell{84.70}
& \neutralcell{92.61}
& \purplecell{78.92}
& \tealcell{82.71}
& \neutralcell{100.71}
& \neutralcell{87.34}
& \neutralcell{163.95} \\

c499-RN1280 & \baselinecell{92.50}
& \tealcell{90.70}
& \tealcell{76.95}
& \purplecell{-8.35}
& \neutralcell{102.05}
& \tealcell{84.05}
& \tealcell{85.05} \\

c499-SL320  & \baselinecell{77.31}
& \neutralcell{78.96}
& \neutralcell{85.30}
& \neutralcell{89.25}
& \neutralcell{89.93}
& \neutralcell{94.40}
& \neutralcell{99.33} \\

c499-SL640  & \baselinecell{83.57}
& \neutralcell{101.86}
& \neutralcell{110.01}
& \purplecell{78.75}
& \neutralcell{99.29}
& \tealcell{84.34}
& \neutralcell{118.16} \\

c499-SL1280 & \baselinecell{92.86}
& \tealcell{77.13}
& \tealcell{68.75}
& \purplecell{56.75}
& \tealcell{52.63}
& \tealcell{85.81}
& \tealcell{91.32} \\

c880-CS320  & \baselinecell{12.75}
& \tealcell{0.74}
& \tealcell{-1.16}
& \purplecell{-2.02}
& \tealcell{-1.35}
& \tealcell{0.67}
& \tealcell{-0.61} \\

c880-CS640  & \baselinecell{11.35}
& \tealcell{-0.76}
& \tealcell{1.38}
& \purplecell{-3.15}
& \tealcell{2.29}
& \tealcell{-0.19}
& \tealcell{-0.48} \\

c880-CS1280 & \baselinecell{11.08}
& \tealcell{0.89}
& \tealcell{-4.60}
& \purplecell{-13.77}
& \tealcell{-1.36}
& \tealcell{2.22}
& \tealcell{1.23} \\

c880-CS2560 & \baselinecell{10.58}
& \tealcell{-2.35}
& \tealcell{3.99}
& \tealcell{-0.52}
& \purplecell{-3.27}
& \tealcell{-1.08}
& \tealcell{1.23} \\

c880-RN320  & \baselinecell{12.37}
& \purplecell{-2.51}
& \tealcell{-1.82}
& \tealcell{-1.32}
& \purplecell{-2.51}
& \tealcell{-0.13}
& \tealcell{-0.75} \\

c880-RN640  & \baselinecell{10.04}
& \tealcell{-0.39}
& \tealcell{-0.24}
& \purplecell{-5.74}
& \tealcell{-0.63}
& \tealcell{-0.63}
& \tealcell{-2.75} \\

c880-RN1280 & \baselinecell{10.13}
& \purplecell{-6.89}
& \tealcell{0.68}
& \tealcell{0.81}
& \tealcell{4.22}
& \tealcell{0.54}
& \tealcell{0.00} \\

c880-RN2560 & \baselinecell{9.31}
& \tealcell{2.40}
& \tealcell{-12.51}
& \purplecell{-15.21}
& \tealcell{-1.82}
& \tealcell{1.29}
& \tealcell{6.11} \\

c880-SL320  & \baselinecell{13.39}
& \tealcell{-2.58}
& \tealcell{1.11}
& \purplecell{-3.07}
& \tealcell{-0.86}
& \tealcell{0.31}
& \tealcell{-2.40} \\

c880-SL640  & \baselinecell{14.15}
& \tealcell{0.53}
& \tealcell{3.42}
& \tealcell{3.66}
& \purplecell{-0.63}
& \tealcell{2.55}
& \tealcell{0.43} \\

c880-SL1280 & \baselinecell{11.20}
& \tealcell{-1.41}
& \tealcell{-3.13}
& \tealcell{-4.58}
& \tealcell{-2.02}
& \purplecell{-5.99}
& \tealcell{-3.77} \\

c880-SL2560 & \baselinecell{9.93}
& \tealcell{-5.34}
& \purplecell{-26.80}
& \tealcell{-26.69}
& \tealcell{-1.14}
& \tealcell{3.67}
& \tealcell{6.11} \\
\bottomrule
\end{tabular}
\textit{Overhead Status: }
\colorbox{gray!20}{\rule{0pt}{4pt}\rule{8pt}{0pt}} baseline\,
\colorbox{yellow!20}{\rule{0pt}{4pt}\rule{8pt}{0pt}} worse\,
\colorbox{green!20}{\rule{0pt}{4pt}\rule{8pt}{0pt}} better\,
\colorbox{green!45}{\rule{0pt}{4pt}\rule{8pt}{0pt}} best
\end{table}

%% file: text/SectionV-III_GNNRE.tex
\subsection{Case Study 3: \textbf{GNN-RE} and AttackGNN}

\begin{figure*}[htbp]
    \centering
    \includegraphics[width=1.0\linewidth]{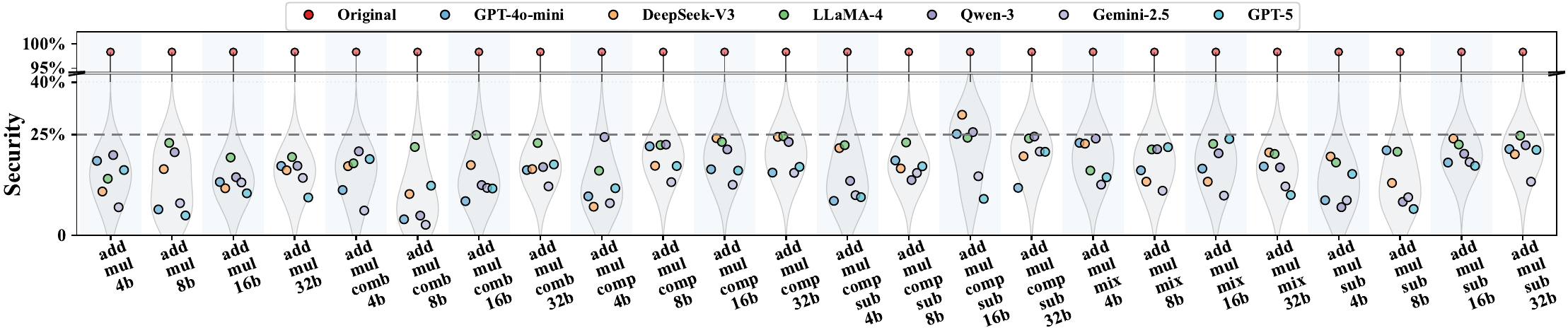} 
    \caption{Security of \textit{NetDeTox} netlists against GNN-RE. Original refers to the baseline design. Scores $\leq 25\%$ indicate successful evasion; the original's score $\approx 100\%$ shows high baseline vulnerability to GNN-RE.}
    \label{fig:gnnre}
    \vspace{-2mm}
\end{figure*}

\noindent

\noindent\textbf{Security Evaluation:} 
For GNN-RE, successful evasion requires classification accuracy $\leq$25\%.
\textit{NetDeTox} reduces GNN-RE's accuracy from 100\% to as low as 4.9\%, which is a 95.1\% drop (Fig.~\ref{fig:gnnre}).
Notably, DeepSeek-V3 and Gemini-2.5 consistently achieve $\leq$25\% across 141/144 cases spanning 4 to 32-bit benchmarks, indicating robust evasion independent of circuit scale.
{AttackGNN also pushes accuracy into the $0$–$25\%$ range, meaning that \textit{NetDeTox} matches its proven evasion effectiveness.}

\noindent
\textbf{Overhead Analysis:} 
%
Tab.~\ref{tab:overhead_b} shows \textit{NetDeTox} consistently outperforms AttackGNN
overheads (54-113\%) across GNN-RE benchmarks. Despite GNN-RE's larger circuit sizes, where area overhead becomes more critical, \textit{NetDeTox} maintains overhead mostly below 40\%, with few cases achieving area
reduction.
LLM selection significantly impacts overhead, with GPT-4o-mini, DeepSeek-V3 and Qwen-3 consistently achieving lower costs, enabling flexible area-security trade-offs.

\input{figs/gnnre_area}

\subsection{Summary for Comparison with AttackGNN}


Across all three tasks, \textit{NetDeTox} consistently breaks OMLA (23/24 pairs at $\sim50\%$), GNN4IP (168/186 at $\leq 0$), and GNN-RE (141/144 at $\leq 25\%$) across six LLM backends.
Beyond robustness, we frequently observe negative overhead, i.e., \textit{NetDeTox} produces more area-efficient circuits than the originals, thanks to efficient exploration of local structures, with scalability consistent across circuit sizes.
Backend-agnostic effectiveness, localized rewriting, and predictable scaling redefine the security–cost trade-off.

\subsection{Case Study 4: \textbf{GNN4IP} and LLMPirate}



\noindent
\textbf{Security Evaluation:}
Recall the previous discussion for \textit{NetDeTox} netlists against GNN4IP (Sec.~\ref{sec:case_GNN4IP}).
{For LLMPirate, most LLMs (CoPilot, GPT-3.5, GPT-4, Claude, and CodeLlama-13B) evade GNN4IP on most netlists,
whereas CodeLlama-7B and Llama3-8B succeed on only 10 and 11 out of 32, respectively, indicating limited flexibility.}

\noindent
\textbf{Overhead Analysis:} 
Our method significantly improves over LLMPirate~\cite{gohil2024llmpirate}. LLMPirate incurs $>200$\% gate-count overheads, whereas \textit{NetDeTox} achieves comparable security with substantially lower overheads
(Tab.~\ref{tab:overhead_model_table}).
Notably, we observe negative overheads, -9.91\% (DeepSeek-V3) and -27.38\% (LLaMA-4), indicating that ours finds much more efficient adversarial structures.
This difference comes from \textit{NetDeTox}'s LLM reasoning about area overhead during planning, selecting gate combinations that minimize cost while altering structures.
In contrast, LLMPirate applies local transformations without area awareness, accumulating uncontrolled overheads.

\input{tabs/overleaf_model_distributions}

\noindent
{\textbf{{Takeaway:}} Context-aware planning enables \textit{NetDeTox} to achieve negative overheads, avoiding LLMPirate's $>200$\% gate inflation through area-conscious transformation selection, all while achieving
	similar evasion performance.}



%% file: figs/gnnre_area.tex
\begin{table}[htbp]
\centering
\caption{Area overhead: AttackGNN vs \textit{NetDeTox} on GNN-RE.}
\label{tab:overhead_b}
\setlength{\tabcolsep}{4pt}
\renewcommand{\arraystretch}{1.0}
\scriptsize
\begin{tabular}{l c|cccccc}
\toprule
\textbf{Case} & \textbf{AG} & \textbf{G4oM} & \textbf{L4} & \textbf{DS-V3} & \textbf{Q3} & \textbf{G2.5} & \textbf{G5} \\
\midrule
add\_mul\_4bit  & \baselinecell{65.06}
& \tealcell{33.42}
& \tealcell{60.00}
& \tealcell{36.96}
& \tealcell{35.44}
& \tealcell{42.28}
& \purplecell{27.59} \\

add\_mul\_8bit  & \baselinecell{75.54}
& \tealcell{15.49}
& \tealcell{28.86}
& \tealcell{49.77}
& \tealcell{14.17}
& \tealcell{17.89}
& \purplecell{13.26} \\

add\_mul\_16bit & \baselinecell{62.76}
& \purplecell{3.51}
& \tealcell{6.35}
& \tealcell{28.14}
& \tealcell{12.44}
& \tealcell{4.97}
& \neutralcell{93.05} \\

add\_mul\_32bit & \baselinecell{72.83}
& \tealcell{3.35}
& \purplecell{1.72}
& \tealcell{10.34}
& \tealcell{4.27}
& \tealcell{12.46}
& \tealcell{9.92} \\

add\_mul\_comb\_4bit & \baselinecell{71.85}
& \purplecell{31.09}
& \tealcell{44.28}
& \tealcell{59.53}
& \tealcell{40.47}
& \tealcell{36.95}
& \tealcell{36.36} \\

add\_mul\_comb\_8bit & \baselinecell{78.95}
& \tealcell{23.66}
& \tealcell{27.80}
& \tealcell{22.75}
& \tealcell{18.67}
& \tealcell{15.57}
& \purplecell{15.33} \\

add\_mul\_comb\_16bit & \baselinecell{54.19}
& \tealcell{8.08}
& \purplecell{-3.19}
& \tealcell{32.14}
& \tealcell{11.07}
& \tealcell{3.38}
& \tealcell{3.78} \\

add\_mul\_comb\_32bit & \baselinecell{73.68}
& \tealcell{6.02}
& \purplecell{0.44}
& \tealcell{4.36}
& \tealcell{6.21}
& \tealcell{18.27}
& \tealcell{4.92} \\

add\_mul\_comp\_4bit  & \baselinecell{61.45}
& \tealcell{32.65}
& \tealcell{14.51}
& \tealcell{37.64}
& \tealcell{28.80}
& \tealcell{27.89}
& \purplecell{32.43} \\

add\_mul\_comp\_8bit  & \baselinecell{73.09}
& \tealcell{9.31}
& \purplecell{5.92}
& \tealcell{14.75}
& \tealcell{6.62}
& \neutralcell{84.18}
& \tealcell{19.70} \\

add\_mul\_comp\_16bit & \baselinecell{61.44}
& \tealcell{6.49}
& \tealcell{10.27}
& \tealcell{12.59}
& \purplecell{3.25}
& \neutralcell{92.43}
& \tealcell{3.45} \\

add\_mul\_comp\_32bit & \baselinecell{71.87}
& \tealcell{24.54}
& \tealcell{13.14}
& \purplecell{4.76}
& \tealcell{4.48}
& \tealcell{24.16}
& \tealcell{16.01} \\

add\_mul\_comp\_sub\_4bit & \baselinecell{60.26}
& \tealcell{19.34}
& \tealcell{17.24}
& \tealcell{25.00}
& \purplecell{13.03}
& \tealcell{17.37}
& \tealcell{18.82} \\

add\_mul\_comp\_sub\_8bit & \baselinecell{70.65}
& \tealcell{3.79}
& \purplecell{3.51}
& \tealcell{6.30}
& \tealcell{8.41}
& \tealcell{6.94}
& \tealcell{5.14} \\

add\_mul\_comp\_sub\_16bit & \baselinecell{62.45}
& \tealcell{9.47}
& \tealcell{2.73}
& \purplecell{1.21}
& \tealcell{3.59}
& \tealcell{31.51}
& \tealcell{7.36} \\

add\_mul\_comp\_sub\_32bit & \baselinecell{70.65}
& \tealcell{8.90}
& \tealcell{11.14}
& \tealcell{10.20}
& \purplecell{4.55}
& \tealcell{25.96}
& \tealcell{5.15} \\

add\_mul\_mix\_4bit & \baselinecell{75.24}
& \tealcell{38.59}
& \neutralcell{80.83}
& \tealcell{41.75}
& \tealcell{44.90}
& \tealcell{54.61}
& \purplecell{33.50} \\

add\_mul\_mix\_8bit & \baselinecell{81.32}
& \purplecell{13.00}
& \tealcell{34.88}
& \tealcell{14.60}
& \tealcell{29.15}
& \neutralcell{91.18}
& \tealcell{14.60} \\

add\_mul\_mix\_16bit & \baselinecell{77.00}
& \tealcell{2.64}
& \tealcell{4.43}
& \tealcell{12.84}
& \tealcell{4.24}
& \tealcell{55.89}
& \purplecell{2.10} \\

add\_mul\_mix\_32bit & \baselinecell{113.02}
& \tealcell{6.99}
& \tealcell{9.05}
& \tealcell{7.30}
& \purplecell{6.90}
& \tealcell{22.86}
& \tealcell{11.71} \\

add\_mul\_sub\_4bit & \baselinecell{61.90}
& \tealcell{25.40}
& \tealcell{23.23}
& \tealcell{18.18}
& \tealcell{21.36}
& \purplecell{17.17}
& \tealcell{26.26} \\

add\_mul\_sub\_8bit & \baselinecell{71.95}
& \tealcell{33.16}
& \purplecell{1.48}
& \tealcell{6.72}
& \tealcell{9.84}
& \tealcell{6.46}
& \tealcell{10.94} \\

add\_mul\_sub\_16bit & \baselinecell{63.63}
& \tealcell{16.89}
& \tealcell{3.35}
& \tealcell{10.93}
& \tealcell{4.29}
& \tealcell{9.10}
& \purplecell{-0.38} \\

add\_mul\_sub\_32bit & \baselinecell{71.69}
& \tealcell{9.45}
& \purplecell{1.71}
& \tealcell{5.67}
& \tealcell{4.22}
& \tealcell{65.51}
& \tealcell{2.63} \\
\bottomrule
\end{tabular}
\textit{Overhead Status: }
\colorbox{gray!20}{\rule{0pt}{4pt}\rule{8pt}{0pt}} baseline\,
\colorbox{yellow!20}{\rule{0pt}{4pt}\rule{8pt}{0pt}} worse\,
\colorbox{green!20}{\rule{0pt}{4pt}\rule{8pt}{0pt}} better\,
\colorbox{green!45}{\rule{0pt}{4pt}\rule{8pt}{0pt}} best
\end{table}

%% file: tabs/overleaf_model_distributions.tex

\begin{table}[htbp]
\centering
\caption{Gate-Count Overheads for GNN4IP.}
\label{tab:overhead_model_table}
\renewcommand{\arraystretch}{1.5}
\footnotesize
\begin{tabular}{@{}m{1.8cm} m{2.2cm} c c c@{}}
\toprule
\textbf{Model} & \textbf{Overhead Dist.} & \textbf{Avg} & \textbf{Min} & \textbf{Std} \\
\midrule

GPT-4o-mini & \centering\includegraphics[width=\linewidth]{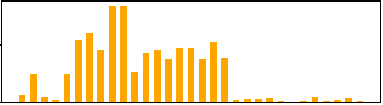}
   & \textbf{103.49\%} & 3.70\% & 1.03 \\
DeepSeek-V3 & \centering\includegraphics[width=\linewidth]{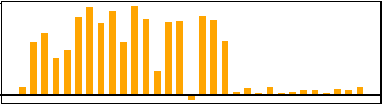}
   & \textbf{82.93\%} & -9.91\% & 0.74 \\
LLaMA-4 & \centering\includegraphics[width=\linewidth]{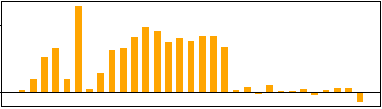}
   & \textbf{81.08\%} & -27.38\% & 0.82 \\
Qwen-3 235B & \centering\includegraphics[width=\linewidth]{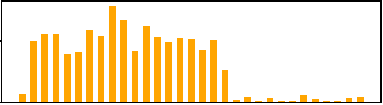}
   & \textbf{124.91\%} & 3.35\% & 1.04 \\
Gemini-2.5 Flash & \centering\includegraphics[width=\linewidth]{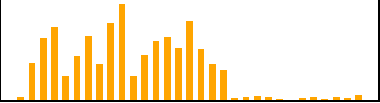}
   & \textbf{124.73\%} & 3.35\% & 1.13 \\
GPT-5 & \centering\includegraphics[width=\linewidth]{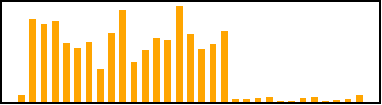}
   & \textbf{138.48\%} & 4.23\% & 1.16 \\

\bottomrule
\end{tabular}
\end{table}

%% file: text/SectionVI-Discussion.tex
\subsection{Ablation Studies}
\subsubsection{RL and LLM Integration }

\begin{figure}[htbp]
    \centering
    \includegraphics[width=1.0\linewidth]{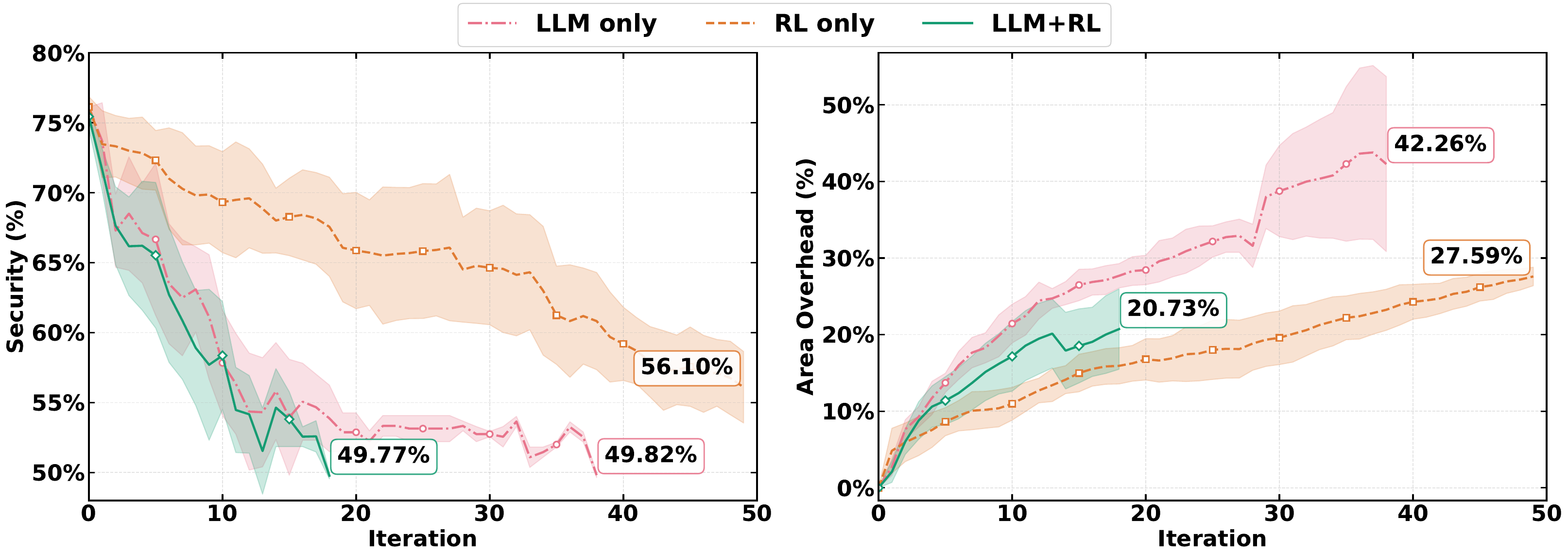} 
    \caption{Ablation case study on OMLA for \texttt{c3540}, averaged over five runs. Shaded areas indicate variance.}
    \label{fig:ablation}
\end{figure}

Here we study the benefits of the proposed hybrid LLM+RL integration within \textit{NetDeTox}.
For example, for OMLA on \texttt{c3540} (Fig.~\ref{fig:ablation}),
we see closely matching attack strength for LLM+RL and LLM-only. However, with about half the area overhead (20.73\% vs 42.26\%) and converging significantly
faster (18 vs 38 iterations), LLM+RL is superior. RL-only stalls at 56.10\% security strength after 49 iterations.
This reveals complementary roles: RL identifies \textit{where} to transform (high-impact locations), while LLM determines \textit{how} to transform (context-aware rewriting).
Neither component alone achieves both effective evasion and area.

\noindent
{\textbf{{Takeaway:}} RL+LLM achieves 50\% less overheads and 50\% faster convergence than LLM-only, while RL-only underperforms for evasion, clearly demonstrating that both components are necessary.}



\subsubsection{RL Queries}

Tab.~\ref{tab:speedup_table} shows that \textit{NetDeTox} reduces RL queries significantly over the RL-only approach in AttackGNN, providing up to $272\times$ speedup.
This finding aligns with faster convergence noted above.
The benefits accrue in all three case studies, i.e., query-efficiency benefits generalize to diverse attack settings.

\noindent
\textbf{Takeaway:}
Beyond security-efficiency trade-offs, \textit{NetDeTox} delivers runtime gains of over 100$\times$, enabling large-scale deployment.

\input{tabs/iteration_speedup}

\subsubsection{LLM Planning Order}
\label{ordering}
\begin{figure}[htbp]
    \centering
    \includegraphics[width=1.0\linewidth]{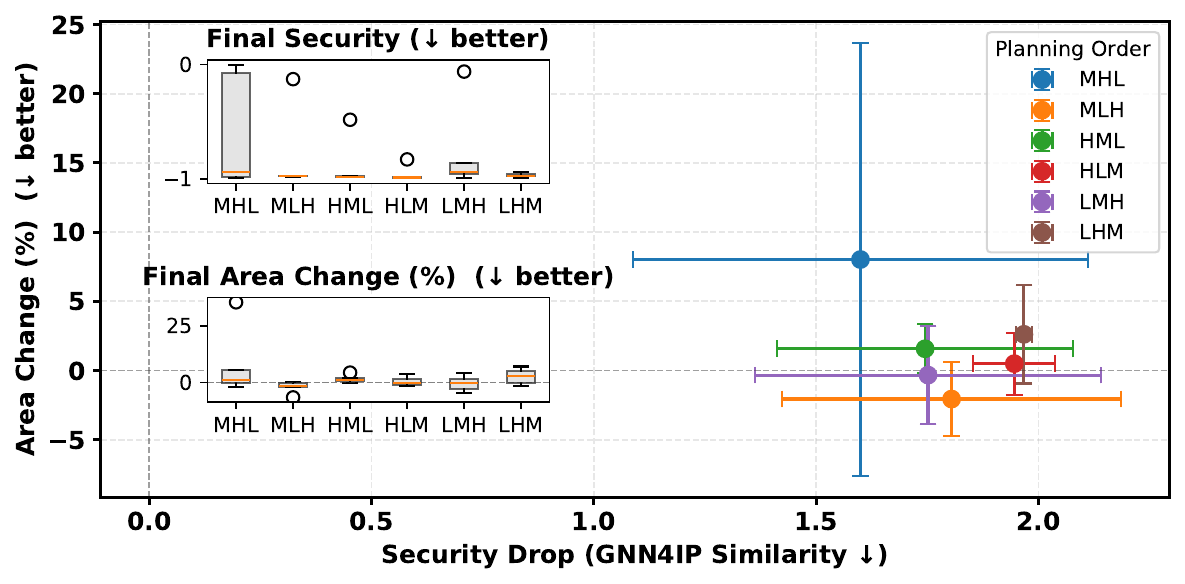} 
    \caption{Efficiency of different planning orders on security-area trade-offs. M denotes subnetlist mapping selection, L denotes gate selection, and H denotes hop size selection.}
    \label{fig:plan_order}
\end{figure}

Here we evaluate whether planning order influences security and area. Without loss of generality, for \texttt{c880-CS2560} on GNN4IP, all possible planning orders are executed five times
(Fig.~\ref{fig:plan_order}),
and the resulting efficiencies are analyzed.
MLH (Mapping–Location–Hop size) and LMH (Location–Mapping–Hop size) outperform other schedules, achieving area reductions efficiently while maintaining the target security. Considering both final area and security, LMH offers the best balance and is adopted as the default for plan generation.

\noindent
\textbf{{Takeaway:}} LMH ordering balances security and area most effectively, showing that planning sequence directly shapes evasion–efficiency trade-offs.

\subsubsection{LLM Planning Tuning}

\begin{figure}[htbp]
    \centering
    \includegraphics[width=1.0\linewidth]{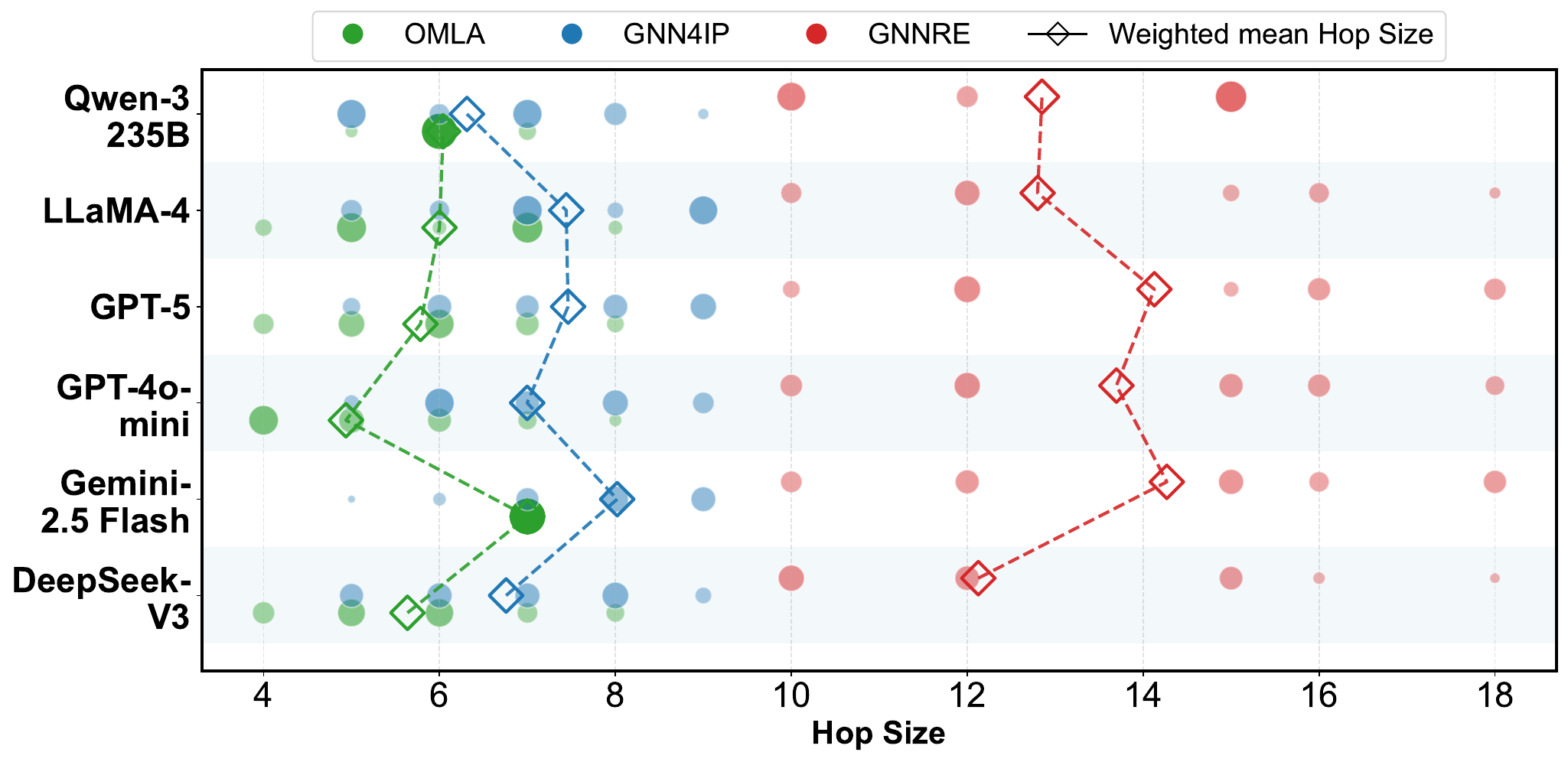} 
    \caption{LLM-preferred hop-size strategies for OMLA, GNN4IP, and GNN-RE. Bubble size indicates relative helpfulness for evasion, while diamonds show weighted mean hop sizes for top-5 choices.}
    \label{fig:gate1}
\end{figure}

\begin{figure}[htbp]
    \centering
    \includegraphics[width=1.0\linewidth]{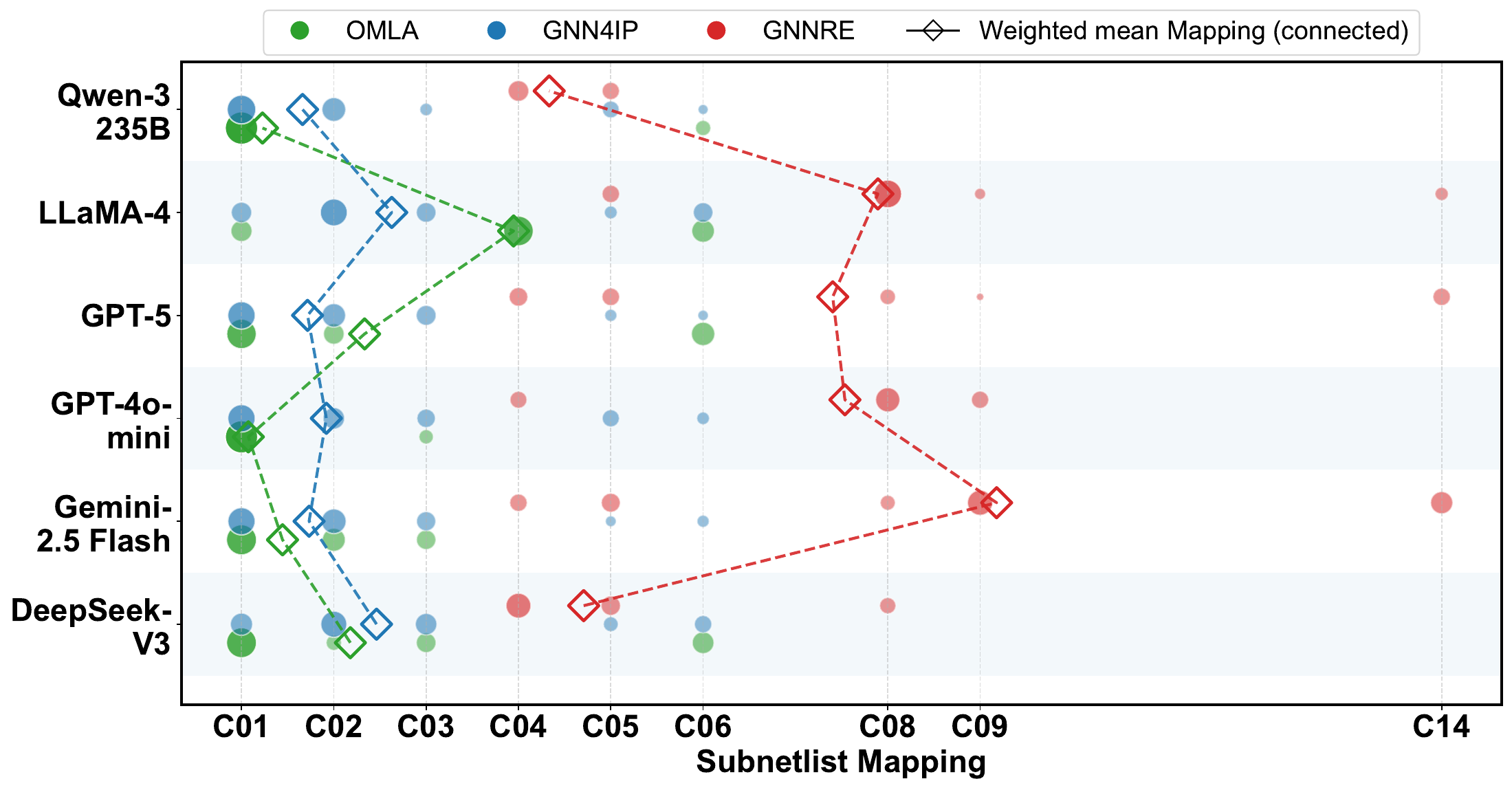} 
    \caption{LLM-preferred subnetlist mapping strategies for OMLA, GNN4IP, and GNN-RE.}
    \label{fig:gate2}
\end{figure}

Figs.~\ref{fig:gate1} and~\ref{fig:gate2} reveal how hop size and subnetlist mapping affect evasion effectiveness across tasks and LLM backends. Hop-size sensitivity varies by task: OMLA achieves best results at smaller
hops ($h=4$-$8$) as it targets localized key-gate structures and was trained on subgraphs with $h \leq 3$, while GNN-RE benefits from larger contexts ($h=12$-$16$) due to its module-level classification.
GNN4IP remains relatively stable across hop sizes, as its graph-level similarity detection is less sensitive to localized transformations.
Mapping selection correlates with hop sizes: smaller hops ($h=4$-$8$) predominantly use mappings C01-C04, while larger hops ($h \geq 12$) favor mappings C08-C09, and other mappings (C06, C07) see limited usage due to
their heavy reliance on XOR/XNOR gates (which introduce area overhead despite effectively altering structure).
No single mapping dominates---optimal choices depend on the target GNN and backend. DeepSeek-V3 and Gemini-2.5 show stable performance across configurations, while GPT-5 exhibits higher variance, suggesting different LLMs have varying sensitivity to transformation granularity.

\noindent
{\textbf{{Takeaway:}} Optimal hop sizes and mappings vary across attacks, and LLM-driven tuning ensures security with low overhead.}

%% file: tabs/iteration_speedup.tex
\begin{table}[htbp]
\centering
\small
\caption{Average RL Queries for AttackGNN vs \textit{NetDeTox}.}
\label{tab:speedup_table}
\renewcommand{\arraystretch}{1.0}
\setlength{\tabcolsep}{3.8pt}
\begin{tabular}{@{}lccccccc@{}}
\toprule
\textbf{Dataset} & \textbf{G4oM} & \textbf{L4} & \textbf{DS-V3} & \textbf{Q3} & \textbf{G2.5} & \textbf{G5} & \textbf{$\downarrow$ Queries} \\
\midrule
OMLA   & 21.75 & \textbf{11.00} & 17.75 & 23.25 & 13.50 & 16.00 & 227.27$\times$ \\
GNN4IP & 14.90 & 9.77  & \textbf{9.16}  & 19.06 & 22.68 & 23.52 & 272.89$\times$ \\
GNN-RE & 39.24 & 57.08 & \textbf{16.88} & 57.08 & 121.25 & 86.25 & 148.15$\times$ \\
\bottomrule
\end{tabular}
\end{table}
\vspace{-5pt}

%% file: text/SectionVII-Conclusion.tex

\section{Conclusion}

We presented \textit{NetDeTox}, a novel framework that combines LLM planning with RL targeting to evade GNN-based security tools through subnetlist rewriting. Focusing on compact subnetlists rather than full circuits or
isolated gates, \textit{NetDeTox} enables efficient, function-preserving edits evading detection.
Compared to prior SOTA, AttackGNN, \textit{NetDeTox} reduces area overheads significantly and achieves comparable (sometimes even better) effectiveness while
evading OMLA, GNN4IP, and GNN-RE, all with
consistent effectiveness over six diverse LLM backends and circuit scales.
Ablation studies show RL and LLM supplement each other well, as only their joint interaction achieves lower overheads and faster convergence than LLM-only and RL-only approaches.
Future work should explore LLM-based synthesis transformations~\cite{li2024circuit}, further evaluation metrics~\cite{10720163}, and other prospects for evasion in other security tasks like side-channel~\cite{brier2004correlation} or
fault-injection~\cite{jain2020atpg} attacks.